\documentclass[%
 aip,
 jcp,
 amsmath,amssymb,
 reprint,%
]{revtex4-2}

\usepackage{graphicx}% Include figure files
\usepackage{dcolumn}% Align table columns on decimal point
\usepackage{bm}% bold math
\usepackage[utf8]{inputenc}
\usepackage[T1]{fontenc}
\usepackage{mathptmx}
\usepackage{etoolbox}

\makeatletter
\def\@email#1#2{%
 \endgroup
 \patchcmd{\titleblock@produce}
  {\frontmatter@RRAPformat}
  {\frontmatter@RRAPformat{\produce@RRAP{*#1\href{mailto:#2}{#2}}}\frontmatter@RRAPformat}
  {}{}
}%
\makeatother
\begin{document}

\preprint{AIP/123-QED}
%(1) Experimental and ab initio (theoretical) studies (study) of the KCs ground state (short-range repulsive) potential above dissociation limit
%(2) Observation and analysis of bound-free transitions to a3?+ and X1?+ states of KCs]
\title{Observation and modelling of bound-free transitions to the $X^1\Sigma^+$ and $a^3\Sigma^+$ states of KCs}
% Force line breaks with \\
\author{V. Krumins}
\author{A. Kruzins}
\author{M. Tamanis}
\author{R. Ferber}
\email{ruvins.ferbers@lu.lv}
\affiliation{Laser Center, Faculty of Physics, Mathematics and Optometry, University of Latvia, 19 Rainis blvd, Riga LV-1586, Latvia}
\author{V. V. Meshkov}
\author{E. A. Pazyuk}
\author{A. V. Stolyarov}
%\email{avstol@phys.chem.msu.ru}
\affiliation{Department of Chemistry, Lomonosov Moscow State University, 119991 Moscow, Leninskie gory 1/3, Russia}
\author{A. Pashov}
\affiliation{Faculty of Physics, Sofia University, 5 James Brouchier blvd., 1164 Sofia, Bulgaria}
%\author{A. Author}
% \altaffiliation[Also at ]{Physics Department, XYZ University.}%Lines break automatically or can be forced with \\
%\author{B. Author}%
%  \homepage{http://www.Second.institution.edu/~Charlie.Author.}
%\affiliation{
%Authors' institution and/or address%\\This line break forced with \textbackslash\textbackslash
%

\date{\today}% It is always \today, today,
             %  but any date may be explicitly specified

\begin{abstract}
The oscillation continuum in laser-induced fluorescence spectra of bound-free $c^3\Sigma^+ \to a^3\Sigma^+$ and (4)$^1\Sigma^+ \to X^1\Sigma^+$ transitions of the KCs molecule were recorded by Fourier-transform spectrometer and modelled under the adiabatic approximation. The required interatomic potentials for ground $a^3\Sigma^+$ and $X^1\Sigma^+$ states were reconstructed in an analytical Chebishev-polynomial-expansion form in the framework of the regularization direct-potential-fit procedure based on the simultaneous consideration of experimental line positions from [R. Ferber et al, Phys. Rev. A, \textbf{80}, 062501 (2009)] and the present \emph{ab initio} calculation of short-range repulsive potential data. It was proved that the  repulsive part over dissociation limit of the derived $a^3\Sigma^+$ potential reproduces the experiment better than the potentials reported in literature. It is also shown that all empirical and semi-empirical potentials available for the $X^1\Sigma^+$ state reproduce the bound-free (4)$^1\Sigma^+ \to X^1\Sigma^+$ spectrum with equal quality in the range of observations.
\end{abstract}

\maketitle

\section{Introduction}\label{sec:Introduction}

The growing interest in increasing the accuracy of potential energy curves of polar alkali diatomics relates to their successful production in ultracold conditions. Fast developing research of cold and ultracold molecular gases revealed their prospective application in various fields of physics and chemistry, see~\cite{Quemener2012} for a review. Among them, the ground-state ultracold KCs molecules have become an object of recent studies~\cite{Grobner2017} due to its large electric dipole moment and their favorable collisional properties~ \cite{Zuchovski2010}. A usual approach is, first, to form the weakly-bound molecules from atom pairs by magneto-association using Feshbach resonances and, second, to transfer them to a deeply-bound rovibrational state by stimulated Raman adiabatic passage (STIRAP)~\cite{Bergmann98}. This procedure requires precise knowledge of the potential energy curves (PECs) of the ground electronic $X^1\Sigma^+$ and $a^3\Sigma^+$ states. The experiment-based singlet and triplet potentials of these states have been determined in~\cite{Ferber2009,Ferber2013} by Fourier-transform spectroscopy in the bound part of interacting $X^1\Sigma^+$ and $a^3\Sigma^+$ states applying a coupled-channel deperturbation procedure. Based on these data the authors of~\cite{Patel2014} performed coupled-channel calculations to predict positions and widths of the interspecies Feshbach resonances in the ultracold K-Cs mixture. These resonances were recently detected in ultracold K-Cs mixture~\cite{Grobner2017}. It has been revealed that the observed positions of the interspecies Feshbach resonances were by about 20 G higher than the respective predictions in~\cite{Ferber2013,Patel2014}. The authors~\cite{Grobner2017} refined the repulsive part of the ground $a^3\Sigma^+$-state PEC located closely to dissociation threshold, using the same form $A+B/R^{n}$, as in~\cite{Ferber2009,Ferber2013}, to match the observed positions of Feshbach resonances. They concluded that it was possible to reproduce the low-energy scattering properties by adjusting only the short-range part of the potential while retaining the intermediate and long-range parts to change the spectroscopic $a^3\Sigma^+$ PEC as little as possible. The resulting analytical PEC allowed to predict with required accuracy the positions of the observed Feshbach resonances~\cite{Grobner2017}. Recently~\cite{Schwarzer21} the $a^3\Sigma^+$ state PECs of KCs, NaCs, and RbCs were constructed employing the Tang-Toennies type model, which is continuous in the entire range of internuclear distances $R$. This model has been used also for KRb triplet $a^3\Sigma^+$ state~\cite{Schwarzer20}. Within this model the repulsive part, including short range, is described in form $(A+BR^{n})exp(-bR-cR^2)$ by five fitting parameters. Though the reported~\cite{Grobner2017,Ferber2009,Ferber2013,Schwarzer21} PECs describe the intermediate bound part of the $a$-state of KCs with high accuracy, a comparision of the short-range parts from different sources reveals substantial differences between them when $R$ is diminishing.

These studies motivated the necessity to obtain a more direct experimental information on the short-range repulsive PEC of the $a^3\Sigma^+$ state, including the part above the dissociation limit. As is known~\cite{Demtroeder2008}, such data may be obtained by recording and analysing the bound-free radiative transitions in laser-induced fluorescence (LIF) from a bound upper-state level that terminate to the lower-state with energies above the dissociation limit. Such transitions produce the modulated continuous ('oscillating') part of a LIF spectrum, which, provided the upper electronic state's PEC is known well enough, is sensitive to the short-range repulsive part of the lower electronic state. The bound-free transitions were observed in (2)$d^3\Sigma^+ -a^3\Sigma^+$ system of the NaK molecule~\cite{Eisel1979} and later developed~\cite{Masters1990} for the  $(1)^3\Pi \rightarrow a^3\Sigma^+$ system to determine the $a^3\Sigma^+$ state repulsive wall above the dissociation limit, which was found to be in reasonable agreement with a prescription  suggested in~\cite{Eryomin1989}. In~\cite{Ferber2000} the repulsive part of $a^3\Sigma^+$ state of NaK was revisited using bound-free $c^3\Sigma^+ \rightarrow a^3\Sigma^+$ transitions. Later the authors of~\cite{McGreehan2011} applied the bound-free part of $(4)^3\Sigma \to a^3\Sigma^+$ emission in NaK to obtain the improved transition dipole moment functions.  In a homonuclear alkali dimer Cs$_2$ LIF to the short-range repulsive part of the $a^3\Sigma^+_u$ state was observed~\cite{Li2007}. More recently, Ashman and co-authors~\cite{Ashman2012} used the recorded bound-free $(5)^3\Pi_0 \rightarrow (1)a^3\Sigma^+$ transitions in NaCs to determine the repulsive wall of the $a^3\Sigma^+$ state and a $(5)^3\Pi_0 \rightarrow a^3\Sigma^+$ relative transition dipole moment function. They demonstrated that applying a simple short-range repulsive wall's shape extrapolated in~\cite{Docenko2006} as $V(R)=A+B/R^{3}$ revealed lack of agreement with the recorded bound-free spectra, both in positions and amplitudes of oscillations. By adjusting the parameters of exponential analytical representation of the repulsive wall potential and the transition dipole moment function the authors of~\cite{Ashman2012} achieved excellent agreement between observed and simulated oscillation structure in wide range of rovibrational levels of the $a^3\Sigma^+$ state in NaCs.

In present study we report on spectroscopic observation and analysis of continuous oscillation structure of bound-free transitions to the repulsive part of both ground singlet $X^1\Sigma^+$ and triplet $a^3\Sigma^+$ states of KCs above the dissociation limit. For $a^3\Sigma^+$ state we applied the $c^3\Sigma^+ \rightarrow a^3\Sigma^+$ bound-free LIF to the repulsive part of the $a^3\Sigma^+$ state potential. The bound-free transitions to the $X$-state were recorded in (4)$^1\Sigma^+ \rightarrow X^1\Sigma^+$ LIF. To model the recorded bound-free spectra we have refitted spectroscopic data from~\cite{Ferber2009} on $a/X$ states in bound region and constructed the respective potentials in a fully analytical Chebychev polynomial expansion (CPE) form, smoothly combined with the short-range part of \emph{ab initio} potentials above dissociation limit, which were calculated in present work. The obtained present short-range part of $a^3\Sigma^+$ state potential as well as the potentials predicted in~\cite{Ferber2009,Ferber2013,Grobner2017,Schwarzer21} were used to calculate the oscillation structure of bound-free transitions, which was compared with the experiment. Modelling of the LIF signal became possible due to recent extended experiment-based data on the $c^3\Sigma^+$-state in~\cite{Szczepkovski2018,Krumins2021}. The (4)$^1\Sigma^+$ state of KCs has been studied in detail in~\cite{Busevica2011, Klincare2012}, hence, its properties are well known. The bound-free part of the (4)$^1\Sigma^+ \rightarrow X^1\Sigma^+$ system was modelled with present $X$-state PEC as well as with PECs from~\cite{Ferber2009, Ferber2013, LeRoy}. The required for simulations spin-allowed (4)$^1\Sigma^+ -X^1\Sigma^+$, $c^3\Sigma^+ - a^3\Sigma^+$ and spin-forbidden (4)$^1\Sigma^+ -a^3\Sigma^+$ transition dipole moment functions are borrowed from~Refs.\cite{Kim2009, Klincare2012, Zaitsevskii2017, Krumins2021}, in which they were obtained in the framework of scalar-relativistic and fully relativistic \emph{ab initio} electronic structure calculations.

\section{Experiment}\label{sec:Experiment}
\subsection{Details}\label{sec:Details}

\begin{figure}
\includegraphics[scale=0.4]{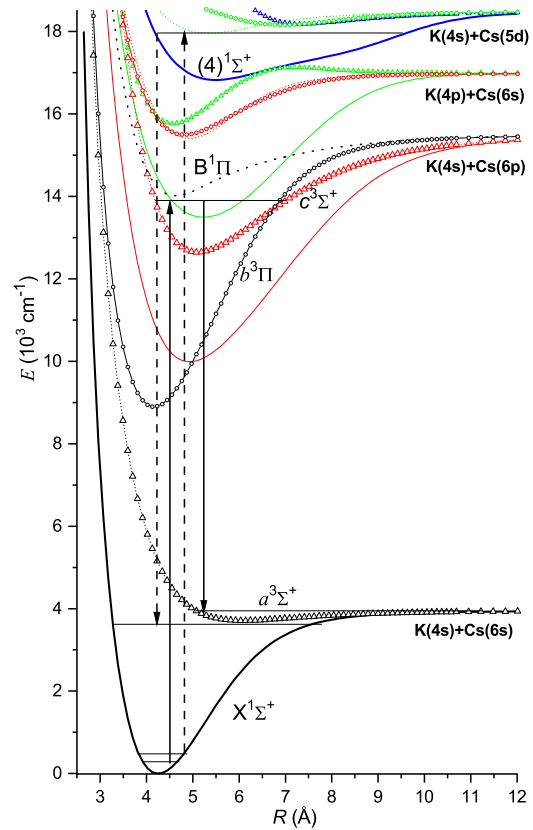}% Here is how to import EPS art
\caption{Excitation and LIF observation schemes used in present study of bound-free transitions in KCs: solid lines for the $a^3\Sigma^+$ state, dotted lines for the $X^1\Sigma^+$ state. Term scheme is borrowed from~\cite{Korek2000}.}\label{fig1:PECs}
\end{figure}

The continuous bound-free transitions in $c^3\Sigma^+\to a^3\Sigma^+$ LIF, see Fig.~\ref{fig1:PECs}, were recorded with the same  experimental set-up as described in our recent study of the $c^3\Sigma^+$  state in KCs~\cite{Krumins2021}. The TiSph laser Solstis/Msqured was used as excitation source. The LIF spectra were recorded by Fourier-transform (FT) spectrometer using an InGaAs diode as detector. KCs molecules were produced in a heat-pipe loaded with K and Cs metals. Buffer gas Ar was added into heat-pipe to prevent condensation of metal vapour on the windows. The main difficulty in the present experiment was to find such excitation transitions, which ensure the most selective excitation of the $c$-state rovibronic levels. Hence, we recorded, at first, a great number of $c^3\Sigma^+\to a^3\Sigma^+$ LIF spectra, see~\cite{Krumins2021}, at high resolution of 0.03 cm$^{-1}$ and analyzed them from the point of view of excitation selectivity of a rovibronic level. Unfortunately, it was not possible to excite only one rovibronic level. At best we have chosen the spectra, in which one very strong progression was dominating while other disturbing progressions  were very weak. To observe the bound-free transitions, LIF spectra from selected suitable rovibronic levels were recorded at diminished spectral resolution of 0.2 cm$^{-1}$ since at high resolution the continuous part of LIF spectra recorded by FT spectrometer is suppressed. These measurements were performed at lowest possible buffer-gas-pressure of about 0.6 mbar at diminished working temperature about 260 $^o$C to minimize the effect of buffer gas induced collisional population transfer between rotational levels  (so called rotational relaxation) in the excited state. The estimated contribution of the latter into obtained bound-free spectra is about 4\%.

In order to record bound-free spectra to the singlet ground $X^1\Sigma^+$ state we used LIF transitions from (4)$^1\Sigma^+$ state, see Fig.~\ref{fig1:PECs}, which was studied in~\cite{Busevica2011}. To excite this state a dye laser Coherent 699-21 with Rhodamin 6G dye was exploited. A photomultiplier tube was used for LIF detection. According to theoretical estimates~\cite{Klincare2012} the bound-free transitions (4)$^1\Sigma^+\to X^31\Sigma^+$ can be observed starting from upper state vibrational levels $v^\prime$ = 44 and higher. Hence, we searched for a selective excitation of high vibrational levels with possibly smaller rotational quantum numbers $J^\prime$. We recorded and analysed LIF spectra at high resolution and chose the ones, in which only one LIF progression was dominating.
For detection of bound-free transitions we recorded preliminary spectra with different spectral resolution, 0.2, 0.5, and 1 cm$^{-1}$. We found that spectral noise diminishes with lower resolution, while the oscillation contrast remains almost unchanged, therefore the bound-free spectra in this case were recorded with resolution 0.5 cm$^{-1}$.

\subsection{The spectra}\label{sec:spectra}

\begin{figure}
\includegraphics[scale=0.4]{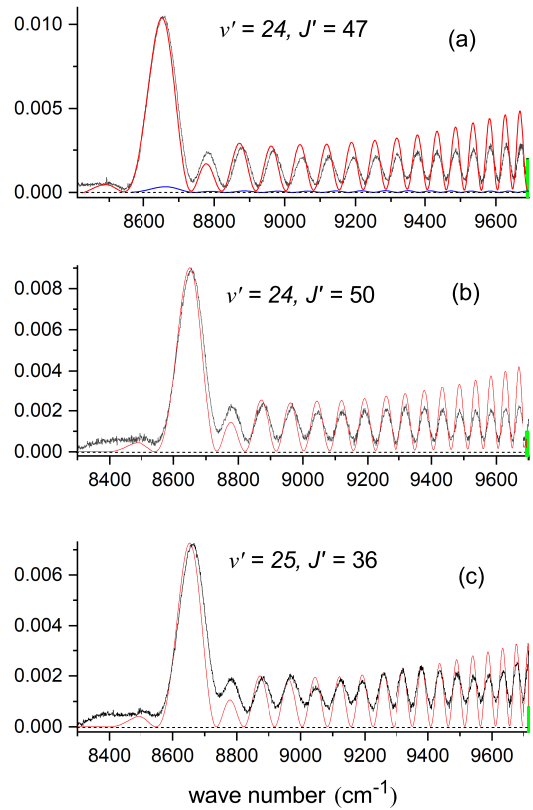}
\caption{Recorded bound-free LIF spectra (black trace) of $c^3\Sigma^+ \to a^3\Sigma^+$ transition from the levels: (a) $v^{\prime}$ = 24, $J^{\prime}$ = 47 $E^{\prime}$ = 13 762.000 cm$^{-1}$, (b) $v^{\prime}$ = 24, $J^{\prime}$ = 50 ($f$-level), $E^{\prime}$ = 13 767.304 cm$^{-1}$; (c) $v^{\prime}$ = 25, $J^{\prime}$ = 36, $E^{\prime}$ = 13 838.207 cm$^{-1}$. Green vertical line at high frequency side marks the end of the discrete spectrum. The bound-free spectra simulated using the present  $a^3\Sigma^+$ PEC and the $c$-state PEC from~\cite{Szczepkovski2018} are shown by red solid lines. The intensity of calculated oscillation spectrum is matched with experiment at the largest peak at 8650 cm$^{-1}$. Blue line in (a) depicts the calculated weak accidentally excited bound-free spectrum from the $v^{\prime}$= 28, $J^{\prime}$ = 130, $E^{\prime}$ = 14194.319 cm$^{-1}$ level. The corresponding integrals over Franck-Condon density,  $\int_{\varepsilon_{min}} |\langle v^{\prime}_c|E_a^{\prime\prime}\rangle_R|^2d\varepsilon$, are equal to 0.7807, 0.7814, and 0.7845 for progressions (a), (b), and (c), respectively.}\label{fig2:c_a_LIF}
\end{figure}

\begin{figure}
\includegraphics[scale=0.4]{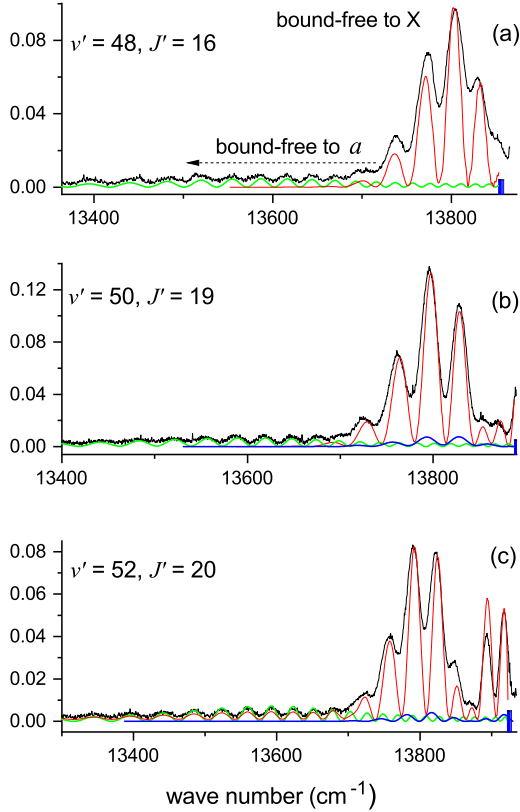}
\caption{Recorded bound-free LIF spectra (black trace) of spin-allowed (4)$^1\Sigma^+ \to X^1\Sigma^+$ and spin-forbidden (4)$^1\Sigma^+ \to a^3\Sigma^+$ transitions. Vertical line at high frequency side marks the end of discrete spectrum of the dominating progression. Solid red lines are the simulated spectra using the present $X^1\Sigma^+$ and $a^3\Sigma^+$ PECs and the (4)$^1\Sigma^+$ PEC from~ \cite{Busevica2011}. The curves are obtained accounting for rotational relaxation and contribution of accidentally excited ($accid$) progressions (blue lines) and normalized to the experiment at the highest peak. Calculated bound-free spectra to the $a^3\Sigma^+$ state normalized to the respective calculated (4)$^1\Sigma^+ \to X^1\Sigma^+$ spectrum are shown by green lines: (a) $v^{\prime}$ = 48, $J^{\prime}$ = 16, $E^{\prime}$ = 17 921.9 cm$^{-1}$; (b) $v^{\prime}$ = 50, $J^{\prime}$ = 19, $E^{\prime}$ = 17 957.1 cm$^{-1}$, $accid$ – $v^{\prime}$ = 58, $J^{\prime}$ = 110, $E^{\prime}$ = 18 237.889 cm$^{-1}$; (c) $v^{\prime}$ = 52, $J^{\prime}$ = 20, $E^{\prime}$ = 17 991.349 cm$^{-1}$, $accid$ – $v^{\prime}$ = 56, $J^{\prime}$ = 16, $E^{\prime}$ = 18 056.672 cm$^{-1}$.  The corresponding integrals over Franck-Condon density, $\int_{\varepsilon_{min}} |\langle v^{\prime}_{(4)^1\Sigma^+}|E_X^{\prime\prime}\rangle_R|^2d\varepsilon$, are equal to 0.3662, 0.3724, and 0.4530 for cases (a), (b), and (c), respectively.}\label{fig3:E_X_LIF}
\end{figure}

The LIF spectra for $c^3\Sigma^+\to a^3\Sigma^+$ transitions were recorded in spectral range from 8000 cm$^{-1}$ to 12000 cm$^{-1}$. Some examples of the recorded bound-free spectra with oscillation structure are presented in Figs.~\ref{fig2:c_a_LIF} and~\ref{fig3:E_X_LIF}, see black traces. For better representation of a bound-free spectrum the discrete part of the spectrum is not shown in these figures. Fig. 2.(a) presents the bound-free part of LIF spectrum from the $c$-state $e$-parity level $v^\prime$ = 24, $J^\prime$ = 47. It represents a sum of two oscillating $P$, $R$ branches with $N_a$ = $J^\prime$ $\pm$ 1, $N_a$ is a rotational quantum number of the $a^3\Sigma^+$. Besides, a weak accidentally excited progression from $c$-state level $v^\prime$ = 28, $J^\prime$ = 130 was also recorded in the discrete spectrum. Fig. 2 (b) shows another example for LIF from the $f$-parity level with $v^\prime$ = 24, $J^\prime$ = 50. The oscillating structure represents a sum of three branches $P^Q$, $Q$, and $R^Q$ with $N_a$ = $J^\prime$+ 2, $N_a$ = $J^\prime$, and $N_a$ = $J^\prime$- 2, with normalized intensity ratios determined from the discrete spectrum as 0.6:1:0.9, respectively. A very weak fragmentary $P$, $R$ progression from the B$^1\Pi$ state level $v^\prime$ = 7, $J^\prime$ = 130, $E^\prime$ = 14 791.666 cm$^{-1}$ was also recorded in high-frequency range. Fig. 2 (c) demonstrates the bound-free spectrum from $e$-level $v^\prime$ = 25, $J^\prime$ = 36. Two weak progressions from levels $v^\prime$ = 33, $J^\prime$ = 111 and $v^\prime$ = 40, $J^\prime$ = 19 were also recorded in the high frequency part. In all three examples the bound-free spectrum of dominating progressions is very weakly contaminated by weak branches from neighbour rotational levels $J^\prime$ due to rotational relaxation, being of about 4\%  from main signal.

Fig.~\ref{fig3:E_X_LIF} shows examples of the recorded bound-free LIF spectra from the (4)$^1\Sigma^+$ state. To focus on the bound-free structure the spectra in Fig. 3 are cut off from both sides. The very end of the discrete part is marked by vertical lines at the upper frequency edge. Along with spin-allowed (4)$^1\Sigma^+ \to X^1\Sigma^+$ transitions the spin-forbidden (4)$^1\Sigma^+ \to a^3\Sigma^+$ transitions are observed as well due to triplet state admixture of the (4)$^1\Sigma^+$ state, see~\cite{Ferber2009, Klincare2012} for details. As can be seen, the continuous bound-free (4)$^1\Sigma^+ \to X^1\Sigma^+$ part markedly differs from the oscillation structure in transitions to the $a$-state in Fig. 2. Namely, the spectra are substantially shorter and contain several peaks with sharply dropping amplitude.  The oscillating structure typical to transitions to the $a^3\Sigma^+$ state is observed at lower frequencies where transition to $X$-state expires. The spectra in Fig.~\ref{fig3:E_X_LIF} (a, b, c) include transitions from (4)$^1\Sigma^+$ state vibrational levels 48, 50, and 52 with rather small $J^\prime$ equal to 16, 19, and 20, respectively.

\section{Modelling}\label{sec:Method}

\subsection{Intensity distribution in the observed LIF spectra}\label{sec:Simulation}

The transition probabilities in the observed (4)$^1\Sigma^+\to X^1\Sigma^+$ and $c^3\Sigma^+\to a^3\Sigma^+$ emission spectra (see Fig.~\ref{fig1:PECs}) have been modelled in the framework of the conventional adiabatic approximation. In particular, this means that a very weak hyperfine interaction between the ground $X^1\Sigma^+$ and $a^3\Sigma^+$ states, as well as the spin-orbit coupling  (SOC) of the excited $c^3\Sigma^+$ state with the nearby $b^3\Pi$ and $B^1\Pi$ states were completely neglected. Though the impact of the SOC effect on the energies of the $c$-state monotonically grows as the vibrational quantum number $v_c^{\prime}$ increases, it does not affect very much the nodal structure of the upper state wavefunctions~\cite{Pupyshev2010} outside the crossing point of $c$- and $b$-states. Regarding the ground states it should be noted that the hyperfine (HFS) $a/X$ mixing is expected to be important only for the highly-excited bound and quasi-bound rovibronic levels located in a vicinity of the dissociation limit.

The relative intensity distribution in the bound-bound (bb) and bound-continuum (bc) parts of the observed (4)$^1\Sigma^+\to X^1\Sigma^+$ and $c^3\Sigma^+\to a^3\Sigma^+$ LIF spectra were evaluated according to the relation
\begin{eqnarray}\label{Itenbc}
I_{ij}^{\textrm{bb}} &\sim& \nu_{ij}^4|\langle v^{\prime}_i|d^{\textrm{ab}}_{ij}|v_j^{\prime\prime}\rangle_R|^2\frac{S_{ij}}{2J^{\prime}+1};\\
I_{ij}^{\textrm{bc}} &\sim& \sum_{J_j^{\prime\prime}} \nu_{ij}^4|\langle v^{\prime}_i|d^{\textrm{ab}}_{ij}|E_j^{\prime\prime}\rangle_R|^2\frac{S_{ij}}{2J^{\prime}+1},
\end{eqnarray}
where $\nu_{ij}=E_i^{\prime}-E_j^{\prime\prime}$ are the transition wave numbers, $S_{ij}(J^{\prime},J^{\prime\prime})$ are the H\"{o}nl-London factors known in the analytical form~\cite{Field2004book}, $i \in [(4)^1\Sigma^+,c^3\Sigma^+]$, and $j\in [X^1\Sigma^+,a^3\Sigma^+]$. The structural (oscillation) continuum $I_{ij}^{\textrm{bc}}$ was calculated on an uniform grid of energy points $E_j$ (with the step $\Delta \varepsilon$ = 0.5 cm$^{-1}$) covering the experimental transition wave numbers, $\nu_{ij}^{expt}$ and satisfied to the complete vibrational basis condition
\begin{eqnarray}\label{sumcond}
\sum_{v^{\prime\prime}=0}^{v^{\prime\prime}_{max}} |\langle v^{\prime}_i|v_j^{\prime\prime}\rangle_R|^2 + \int_{\varepsilon_{min}} |\langle v^{\prime}_i|E_j^{\prime\prime}\rangle_R|^2d\varepsilon=1 ,
\end{eqnarray}
where the minimal $\varepsilon_{min}$-value corresponds to the energy of the last found (highest) quasi-bound level, $v^{\prime\prime}_{max}$.

Hereafter, $d^{\textrm{ab}}_{ij}(R)$ are the electronic transition dipole moments which were borrowed for spin-allowed (4)$^1\Sigma^+\to X^1\Sigma^+$, $c^3\Sigma^+\to a^3\Sigma^+$ and spin-forbidden (4)$^1\Sigma^+\to a^3\Sigma^+$ transitions from electronic structure calculations performed in the framework of both scalar-relativistic~\cite{Kim2009, Klincare2012} and fully relativistic~\cite{Zaitsevskii2017, Krumins2021} approximations. The achieved results have demonstrated that the absolute accuracy of the state-of-art \emph{ab initio} $d^{\textrm{ab}}_{ij}(R)$ functions derived, at least, for strong  spin-allowed transitions becomes comparable with the uncertainty of intensity measurements.

The eigenvalues and eigenfunctions required for the particular (bound, quasi-bound and continuum) rovibronic levels of upper and lower electronic states were obtained from the numerical solution of the 1D radial equation~\cite{Abarenov90}:
\begin{eqnarray}\label{radial}
\left(\frac{\hbar^2 d^2}{2\mu dR^2} + U^{ad}(R) + \frac{\hbar^2 J(J+1)}{2\mu R^2}- E_i^{calc}\right)|v_i\rangle=0
\end{eqnarray}
with the corresponding adiabatic PEC, $U^{ad}(R)$. The ro-vibrational wavefunctions $|v_i\rangle$ belonging to bound and quasi-bound levels were conventionally normalized as $|\langle v_i|v_i\rangle_R|^2=1$ while the energy-normalized continuum wavefunctions $|E_j^{\prime\prime}\rangle$ were obtained by scaling a numerically propagate wavefunctions on the relevant amplitudes of their analytical second-order semi-classical counterparts~\cite{Abarenov92} taken in the vicinity of equilibrium distance $R_e$ of the corresponding $U^{ad}(R)$ potential.

The adiabatic potential of the upper (4)$^1\Sigma^+$ state determined in the fully analytical Chebychev-polynomial-expansion (CPE) form was taken from~\cite{Busevica2011}, while the point-wise inverted-perturbation-approach (IPA) potential~\cite{Szczepkovski2018} was implemented for the excited $c^3\Sigma^+$ state.

\subsection{Semi-empirical potentials for $X^1\Sigma^+$ and $a^3\Sigma^+$ states}
%\subsection{Analytical Semi-empirical interatomic potential construction for  for $X^1\Sigma^+$ and $a^3\Sigma^+$ states}

Interatomic $U^{ad}(R)$ PECs for both singlet and triplet ground states were defined on the semi-interval $R\in [R_{min},+\infty)$ in the CPE form~\cite{Busevica2011}:
\begin{eqnarray}\label{CPE}
U_{CPE}(R) = T_{dis} - \frac{\sum_{k=0}^{m}c_k T_k(y_p)}{1+\left(R/R_{ref}\right)^n},
\end{eqnarray}
where $T_k(y)$ are the Chebyshev polynomials of the first kind depending on the reduced radial variable $y_p(R)\in [-1,1]$:
\begin{eqnarray}\label{Chebyshevy}
y_p(R)=\frac{R^p-R_{ref}^p}{R^p+R_{ref}^p-2R_{min}^p},
\end{eqnarray}
in which $p$ is a small positive integer and $R_{ref}>R_{min}\geq 0$ is a reference distance.

The CPE coefficients $c_k$ were obtained in the framework of the regularization direct-potential-fit (DPF) procedure based on the iterative non-linear minimization of the sum of squares:
\begin{eqnarray}\label{chisquared}
\sum_{l=1}^{N_{expt}}\left (\frac{\nu^{expt}_l-\nu^{calc}_l}{\sigma^{expt}_l}\right )^2 + w\sum_{l=1}^{N_{ab}}\left (\frac{U^{CBS}(R_l)-U^{CPE}(R_l)}{\sigma^{CBS}(R_l)}\right)^2,
\end{eqnarray}
where the weight factor $w$ was selected to provide the proper balance between experimental and \emph{ab initio} data contributions into the total sum. $N_{expt}$ and $N_{ab}$ are the numbers of experimental lines and \emph{ab initio} PEC points involved in the sum, respectively. The input \emph{ab initio} data mainly determine the repulsive part of the $U^{CPE}(R)$ potential (above dissociation limit) while and the experimental data provide an accurate reconstruction of the well-bound (below dissociation limit) part of the resulting semi-empirical potential.

The CPE analytical form (\ref{CPE}) possesses the correct long range asymptotic behavior
\begin{eqnarray}\label{LR}
U_{CPE}(R\to +\infty) = T_{dis} - \sum_{n}\frac{C_n}{R^n},
\end{eqnarray}
where the leading dispersion coefficients $C_n$ are explicit functions of CPE parameters~\cite{Busevica2011}:
\begin{eqnarray}\label{C6}
C_n= R_{ref}^n\sum_{k=0}^{m}c_k,
\end{eqnarray}
\begin{eqnarray}\label{C8C10}
C_{n+p}= 2(R_{min}^p-R_{ref}^p)\sum_{k=0}^{m}k^2c_k .
\end{eqnarray}
Hence, the fitting coefficients $c_k$ were constrained owing to the relations (\ref{C6}) - (\ref{C8C10}) in order to fit the $C_6$, $C_8$, and $C_{10}$ values available from the relevant \emph{ab initio} atomic calculations~\cite{Derevianko2001, Porsev2003}.

In expression (\ref{chisquared}), $\nu^{expt}_l$ are the experimental wave numbers corresponding to the bound-bound part of the LIF spectra below the dissociation limit. These values, together with their uncertainties $\sigma^{expt}_l$ were borrowed from the Supplemented Material of Ref.~\cite{Ferber2009}. All mutually HFS-perturbed levels of both singlet and triplet states closely-lying to their common dissociation threshold were excluded from the present fit. The calculated transition energy $\nu^{calc}_l\equiv \nu^{calc}_{ij}$ is $E^{\prime}_i-E_j^{\prime\prime}$, where $E^{\prime}_i$ and $E_j^{\prime\prime}$ are rovibronic energies of the upper and lower electronic states, respectively. The $E_i^{\prime}$-values are considered to be the adjusted fitting parameters while $E_j^{\prime\prime}\equiv E^{calc}_i$ are the eigenvalues of the radial equation (\ref{radial}) solved with the present CPE potential (\ref{CPE}).

The $U^{CBS}(R)$ PEC involved in Eq.(\ref{chisquared}) was obtained by an extrapolation to the complete basis set (CBS) electronic energy $U^{CCSD(T)}$, which was calculated by means of the spin-restricted close-shell coupled-cluster approach with perturbative treatment of triple excitation (CCSD(T)) and the pseudo-potentials of the \emph{aug-cc-pwCVnZ-PP} (with $n$ = 4 and 5) quality for both K and Cs atoms~\cite{Hill2017, Lim2005}. The electronic structure calculation was conducted by means of the MOLPRO 2010.1 package~\cite{molpro}. All 18 (2 valence + 16 sub-valence) electrons of the molecule were explicitly correlated. For the CBS extrapolation the two-point formula was adopted:
\begin{eqnarray}\label{cbs}
U^{CCSD(T)}(n)=U^{CBS}+\frac{A}{n^\alpha} ,
\end{eqnarray}
where the degree parameter $\alpha$ was adjusted, by the substitution of $U^{CBS}$ for $U^{emp}$ in Eq.(\ref{cbs}). This allowed to achieve that the resulting extrapolated energy perfectly fits the repulsive part of the corresponding empirical PECs $U^{emp}(R)$, which have been already well-established in the well-bound region (below the dissociation limit) in the ground states studies~\cite{Ferber2009, Ferber2013}. Then, the linear extrapolation of the derived $\alpha(R)$ functions to smaller $R$-values (for $R\leq R_{min}$, where $U^{CPE}(R_{min})=T_{dis}$) was performed and the extrapolated $\alpha$-values were applied for a direct estimate, by Eq.(\ref{cbs}), of the repulsive limb of the $U^{CBS}(R\leq R_{min})$ PEC above the dissociation limit. The uncertainties $\sigma^{CBS}$ in the extrapolated PEC $U^{CBS}$ were roughly estimated as a difference between the $U^{CBS}$ values obtained with the fixed $\alpha$-values taken as $\alpha$ = 3 and 4. These point-wise $U^{CBS}(R_l)$ and $\sigma^{CBS}(R_l)$ functions were finally involved in the sum (\ref{chisquared}).

\section{Results}
\subsection{Potential energy curves}\label{sec:A}

\begin{figure}
\includegraphics[scale=0.4]{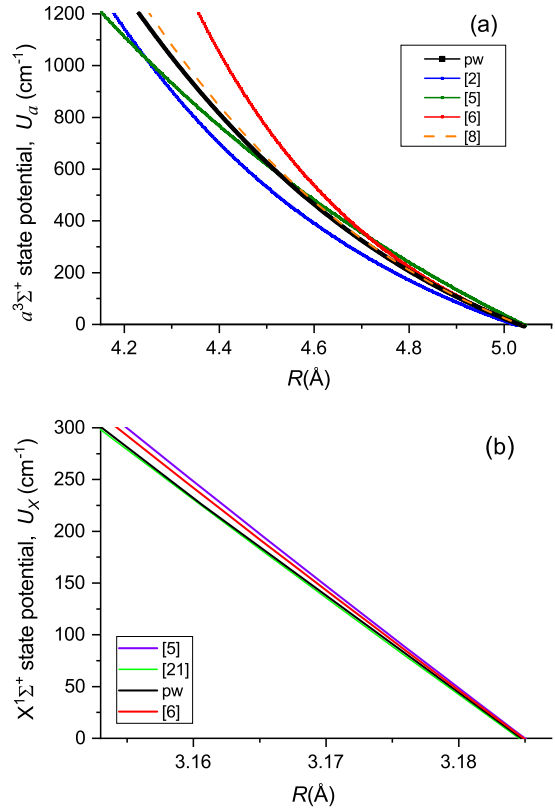}
\caption{Comparison of short-range semi-empirical potentials of the $a^3\Sigma^+$ (a) and $X^1\Sigma^+$ (b) states of KCs above dissociation limit taken from different sources; pw - present work.}\label{fig4a:aPECs-dif}
\end{figure}

The resulting parameters of obtained semi-empirical CPE interatomic potentials of the $a^3\Sigma^+$ and $X^1\Sigma^+$ states are presented in Table~\ref{Chebcoeff}. These fully analytical PECs were used to describe the recorded bound-free LIF spectra to $a^3\Sigma^+$ and $X^1\Sigma^+$ states. The short-range part of $a^3\Sigma^+$ and $X^1\Sigma^+$ potentials over dissociation limit in the energy range covered by present experiment is presented in Fig. \ref{fig4a:aPECs-dif} along with the previously constructed empirical potentials borrowed from previous works. As can be seen in Fig. \ref{fig4a:aPECs-dif} (a), the present $a$-state potential (pw) is very close to the analytical semi-empirical potential recently obtained in Ref.~\cite{Schwarzer21}; they start to differ slightly only at energies about 600 cm$^{-1}$. According to Fig.\ref{fig4a:aPECs-dif} (b) the present $X$-state potential (pw) defined in the CPE form practically coincides with its Morse/Long-Range (MLR) counterpart~\cite{LeRoy}. Robert LeRoy has used to build the MLR potential (given in the Supplementary Material~\cite{EPAPS}) exactly the same LIF line positions~\cite{Ferber2009} as we did but he did not involve in the DPF procedure any \emph{ab initio} data. The slope of both CPE and MLR potentials is slightly smaller than the slope of potentials constructed in the hybrid, so-called Hannover's, form in Refs~\cite{Ferber2009, Ferber2013}.

\subsection{The $a^3\Sigma^+$ state}\label{sec:a3Sig}

Let us analyze how the repulsive part of the present semi-empirical $a$-state CPE potential (\ref{CPE}) describes the experimental bound-free $c-a$ spectra. The calculated bound-free spectra are depicted in Fig.~\ref{fig2:c_a_LIF} by solid red curves, which amplitudes are matched with experiment at the largest peak located at about 8650 cm$^{-1}$; note that the calculated oscillation spectra are averaged over rotational branches. The curves include weak rotational relaxation, which contributes by about 4\% to the total oscillation spectrum. According to our analysis the rotational relaxation does not cause any shifting of peak positions leading only to a slight broadening of oscillation structure. The simulated signal in Fig.\ref{fig2:c_a_LIF} (a) contains also a contribution from an accidentally excited level $v^\prime$ = 28, $J^{\prime}$ = 130 of the $c$-state with energy $E^{\prime}$= 14 194.284 cm$^{-1}$, see the small-amplitude blue curve. The simulated curve in Fig.\ref{fig2:c_a_LIF} (b) is averaged over three rotational branches $N_a$  = $J^\prime$ + 2, $N_a$  = $J^\prime$, and $N_a$  = $J^\prime$- 2, with normalized intensity ratios 0.6:1:0.9, respectively. We have not included here the 5\% contribution from the $B^1\Pi$ state level $v^\prime$ = 7, $J^{\prime}$= 130, $E^{\prime}_B$ = 14 791.666 cm$^{-1}$ since a correct accounting would need the knowledge of non-adiabatic multi-channel structure of upper state wavefunction. For the same reason we did not include in Fig.\ref{fig2:c_a_LIF} (c) the 7\% contribution from the $c$-state level $v^\prime$ = 33, $J^{\prime}$ = 111, $E^{\prime}$ = 14 314.418 cm$^{-1}$. As can be seen in these examples the calculated intensities describe satisfactory the oscillation peaks’ positions. However, it is clearly seen that oscillations’ amplitude in the higher frequency range markedly exceeds the experimental one; the related discussion is postponed to Sec.~\ref{sec:conclusions}.

\begin{figure}
\includegraphics[scale=0.4]{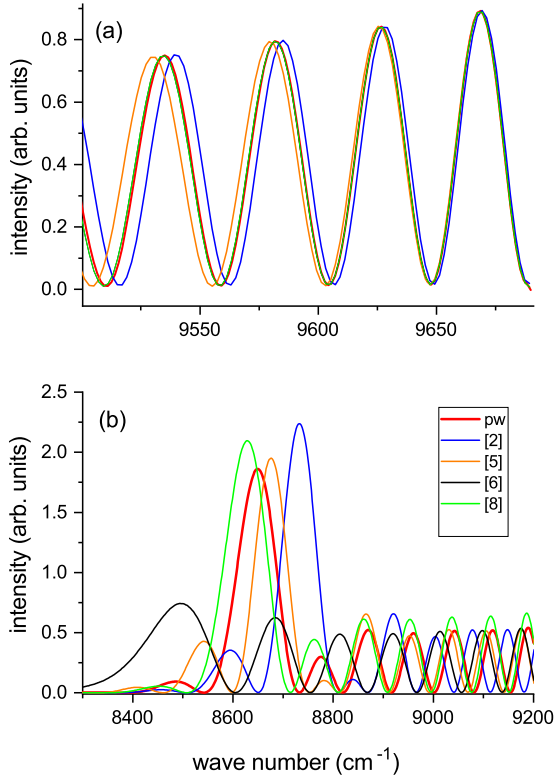}
\caption{Comparison of calculated bound-free $c^3\Sigma^+ \to a^3\Sigma^+$ spectrum from the upper level $v^{\prime}$ = 24, $J^{\prime}$ = 47 using different $a^3\Sigma^+$ semi-empirical PECs: red – present, blue–~\cite{Grobner2017}, green–~\cite{Schwarzer21}, orange–~\cite{Ferber2009}, and black–~\cite{Ferber2013}. (a) first four oscillations of the continuous spectrum just over dissociation limit; (b) the lower frequency part of continuous spectrum.}\label{fig4:aPECs}
\end{figure}

It is of interest to compare the calculated oscillation patterns obtained applying the $a^3\Sigma^+$ state short-range potentials above dissociation limit from different sources. Fig.~\ref{fig4:aPECs} presents an example of simulated bound-free part of the $c-a$ spectrum from $v^\prime$ = 24, $J^{\prime}$ = 47 level; the pattern obtained with present CPE potential is depicted with by red line. For visibility the simulated oscillations are divided in two parts. Fig.~\ref{fig4:aPECs} (a) includes the frequency range just above the $a$-state dissociation limit where the first oscillation starts. It is obvious that meaningful differences between the signals obtained by the present potential and by PECs borrowed from the literature appear even at second oscillation for Ref.~\citep{Ferber2009}, and for the third oscillation for Ref.~\citep{Grobner2017} and the differences are increasing with oscillation number. The differences become overwhelming at low frequency part below 9000 cm$^{-1}$, see Fig.~\ref{fig4:aPECs} (b).

To compare quantitatively the potentials agreement with the experiment, we have determined the differences between experimental and calculated positions of oscillations peaks. These differences were transformed from frequency scale to energy differences $\Delta E=E^{expt} - E^{calc}$ with respect to energy $\varepsilon$, Eq.(\ref{sumcond}), above dissociation limit for a measured progression, see green vertical lines at high frequency side in Fig.~\ref{fig2:c_a_LIF}. The differences are represented in Fig.~\ref{fig5:Difmax} as dependent on $\varepsilon$. The differences $\Delta E(\varepsilon)$ obtained with the present CPE potential are depicted by full circles. The respective values are distributed within $\pm$ 1.5 cm$^{-1}$ corridor in the range up to 400 cm$^{-1}$, then they start to decline slightly from the experiment. For instance, for progression with $v^\prime$ = 25, $J^{\prime}$ = 36 the difference reaches the value of 10 cm$^{-1}$ at energy of 1050 cm$^{-1}$ over dissociation limit. The differences $\Delta E(\varepsilon)$ obtained with the alternative forms of potential taken from Refs~\cite{Grobner2017, Ferber2009, Ferber2013, Schwarzer21} are depicted by empty squares for progression with $v^\prime$ = 24, $J^{\prime}$ = 47. As it can be seen, the PECs~\cite{Grobner2017, Ferber2009, Ferber2013} yield the respective values, which start to decline from the experiment very fast. Note that the last point at 1000 cm$^{-1}$ for PEC taken from Ref.~\cite{Ferber2013} is out of scale by 158 cm$^{-1}$. The  $\Delta E(\varepsilon)$ values obtained with the semi-empirical PEC recently constructed in Ref.~\cite{Schwarzer21} are much closer to the one obtained by the present semi-empirical potential, however, they still start to decline at energies over 500 cm$^{-1}$.

\subsection{The $X^1\Sigma^+$ state}\label{sec:X1Sig}

The simulations of observed bound-free spectra to the ground singlet $X$-state are presented in Fig.~\ref{fig3:E_X_LIF} using the present $X$-state CPE potential and the (4)$^1\Sigma^+$ CPE PEC borrowed from Ref.~\cite{Busevica2011}. The simulated signal accounts for the weaker bound-free spectrum to the triplet $a$-state as well. The intensities calculated for the spin-forbidden (4)$^1\Sigma^+ - a^3\Sigma^+$ transition (depicted by a green line) were normalized to the theoretical intensities of the spin-allowed (4)$^1\Sigma^+ \to X^1\Sigma^+$ transition (red line). The simulations account for rotational relaxation, which contributes overall of about 10\% to the bound-free spectrum. The simulations have demonstrated that this effect causes some broadening while it does not shift the oscillation peak positions. The simulated signals in Fig.~\ref{fig3:E_X_LIF} (b), (c) account for a contribution from accidentally excited levels depicted by small-amplitude blue curves. As can be seen, the simulated spectra describe satisfactorily the recorded spectra. The tiny non-systematic deviations of peaks maxima are most probably caused by additional background ignored in the modeling. The simulation proves the appearance of additional peaks, which are becoming more pronounced with increasing upper state vibrational quantum number, $v^\prime$, see the signals in Fig.~\ref{fig3:E_X_LIF}, (b) and (c).

It should be mentioned that for $v^\prime$ = 52 we also have performed testing calculations of the expected (4)$^1\Sigma^+ - X^1\Sigma^+$ bound-free spectra using different empirical potentials which are available for the $X^1\Sigma^+$ state from previous works, namely~\cite{Ferber2009, Ferber2013, LeRoy}. The oscillation pattern was practically identical for all four exploited PECs; the corresponding figure is presented in Supplemented Material.

%******************************** Table I ********************************
\begin{table}%[H] add [H] placement to break table across pages
\caption{The resulting mass-invariant parameters of the CPE
potential (\ref{CPE}) defined for the $X^1\Sigma^+$ and $a^3\Sigma^+$ states of KCs
on the semi-interval $[R_{min},+\infty)$. The dissociation energy $T_{dis}=4069.076$ (in cm$^{-1}$)
referring to the minimum of the ground $X^1\Sigma^+$ state. The Chebyshev expansion coefficients $c_k$ are given
in cm$^{-1}$ while the fixed $R_{min}$ and $R_{ref}$ parameters in \AA.
The common $n$ = 6 and $p$ = 2 parameters were also fixed for both states.
The FORTRAN subroutine generating the present CPE potentials for both $X^1\Sigma^+$ and $a^3\Sigma^+$ states of KCs
is provided in the Supplementary Material~\cite{EPAPS}.} \label{Chebcoeff}
%\begin{ruledtabular}
\begin{tabular}{crr}
\hline\hline
 & $X^1\Sigma^+$  & $a^3\Sigma^+$\\
 \hline
 $c_0$   &  -373.624 &  1557.978  \\
 $c_1$   & -5391.769 & -3532.255  \\
 $c_2$   &  6442.491 &  2448.655  \\
 $c_3$   & -2248.684 & -1132.530  \\
 $c_4$   &   307.482 &   376.473  \\
 $c_5$   &  -679.865 &   -82.734  \\
 $c_6$   &   295.480 &    -7.879  \\
 $c_7$   &    30.805 &    16.981  \\
 $c_8$   &    87.482 &   -15.873  \\
 $c_9$   &   -42.874 &    20.270  \\
 $c_{10}$&   -46.075 &   -16.752  \\
 $c_{11}$&    -0.775 &     8.659  \\
 $c_{12}$&    17.062 &    -4.080  \\
 $c_{13}$&     6.304 &     1.238  \\
 $c_{14}$&    10.069 &     0.049  \\
 $c_{15}$&    -4.478 &   \\
 $c_{16}$&     5.343 &   \\
 $c_{17}$&    -3.195 &   \\
 $c_{18}$&     1.150 &   \\
 $c_{19}$&    -2.588 &   \\
 $c_{20}$&    -0.434 &   \\
 $c_{21}$&    -0.685 &   \\
  \hline
 $R_{min}$   & 2.5  & 3.0\\
 $R_{ref}$   & 5.0  & 6.4\\
 \hline
\end{tabular}
\end{table}

\section{Concluding remarks}\label{sec:conclusions}

\begin{figure}
\includegraphics[scale=0.4]{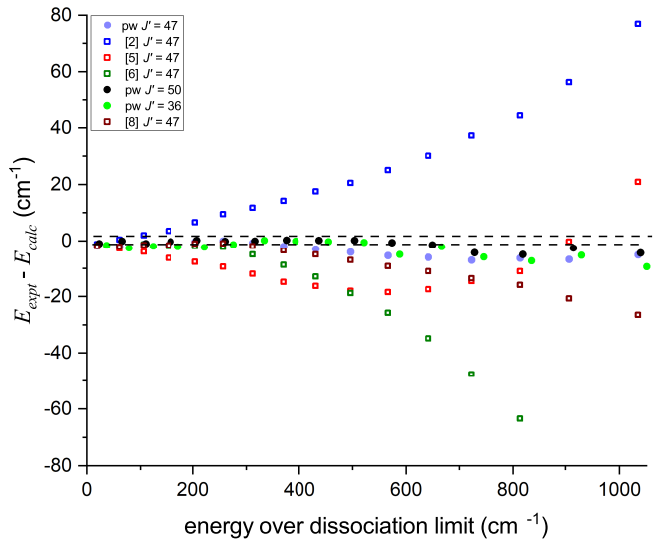}
\caption{Energy differences $\Delta E(\varepsilon)=E^{expt} - E^{calc}$  between experimental energies $E^{expt}$ for measured LIF progressions and the respective energies $E^{calc}$ calculated using the $a^3\Sigma^+$ PECs taken from different works as dependent on energy over dissociation limit, $\varepsilon$. Dashed horizontal lines depict experimental uncertainty as $\pm$ 1.5 cm$^{-1}$ limits.}\label{fig5:Difmax}
\end{figure}

\begin{figure}
\includegraphics[scale=0.4]{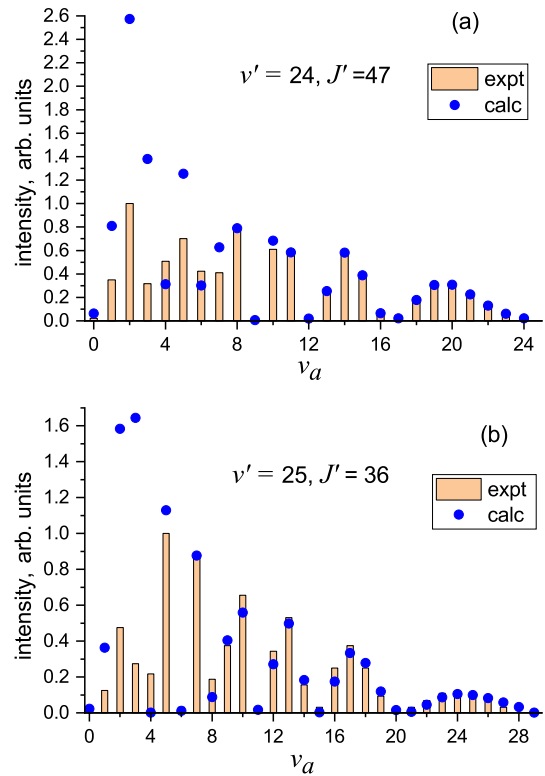}
\caption{Relative intensity distribution in $c^3\Sigma^+ \to a^3\Sigma^+$ bound - bound LIF progression coming from upper levels (a) $v^{\prime}$ = 24, $J^{\prime}$ = 47 and (b) $v^{\prime}$ = 25, $J^{\prime}$= 36; columns - experiment, full circles - calculation. The calculated intensities are matched with experiment at $v_a$ = 8 and $v_a$ = 7 bands for the progressions of (a) and (b), respectively.}\label{fig7:intens}
\end{figure}

Both Fig.~\ref{fig2:c_a_LIF} and Fig.~\ref{fig5:Difmax} demonstrate that the conventional adiabatic approximation used in present work describes rather well the observed oscillatory structure of the bound-free $c^3\Sigma^+ \to a^3\Sigma^+$ LIF spectrum. However, some discrepancies should be mentioned. First, the calculated oscillations amplitude in the beginning part of the bound-free spectrum exceeds the experimental one, see Fig.~\ref{fig2:c_a_LIF}. Second, the small systematic difference between present calculated and experimental energies of the short-range part of repulsive potential is seen in Fig.~\ref{fig5:Difmax}. A possible reason, which can influence the current results is the fact that the $c^3\Sigma^+$ state is  perturbed by the close-lying (see Fig.~\ref{fig1:PECs}) singlet $B^1\Pi$ state and also by the triplet $b^3\Pi$ state, forming the so-called spin-orbit $B\sim b\sim c$ complex, see~\cite{Krumins2021} and references therein. Accordingly, the $c$-state wavefunction should be affected by these perturbations, hence, the usage of upper state adiabatic vibrational wavefunction strictly speaking is not correct. For example, Fig.~\ref{fig7:intens} illustrates a relative intensity distribution measured for the bound-bound part of the $c \to a$ LIF progression from the levels $v^\prime$ = 25, $J^{\prime}$ = 36 and $v^\prime$ = 24, $J^{\prime}$ = 47. As can be seen, though the intensities calculated under adiabatic approximation nicely reproduce the experiment for transitions to higher $v_a$ above the matching vibrational level, they completely fail for transitions to lower levels. The apparent reason for the observed discrepancy is the spin-orbit interaction of $c$- and $b$-states in the vicinity of their crossing point (see Fig.~\ref{fig1:PECs}).  Therefore, a comprehensive deperturbation analysis of the spin-orbit coupled $B\sim b\sim c$ complex is indispensably required to refine a nodal structure of vibrational multi-channel wavefunction of the upper $c$-state. This extremely challenging and complicated task is still to be realized. The respective work is currently in progress.

As is shown by Fig.~\ref{fig5:Difmax}, though the Feshbach resonances are a very sensitive probe of the quality of the $a$-state PEC in the vicinity of dissociation limit (see, for instance, Ref.~\cite{Grobner2017}) they do not guarantee the correct PEC behaviour for higher energies above the dissociation limit. Here, the informations based on bound-free transitions experiments is useful.

\section*{Acknowledgements}
We are indebted to Michael Schwarzer and Jan Peter Toennies for providing their $a^3\Sigma^+$ state potential in the appropriated format and to Ilze Klincare for numerous helpful pieces of advice. Riga team acknowledges the support from the Latvian Council of Science, project No. lzp-2020/2-0215: "Interatomic potentials of alkali atom pairs in a wide range of internuclear distances" and from the University of Latvia Base Funding No A5-AZ27. Moscow team is grateful for the support by the Russian government budget (section 0110), projects No.121031300173-2 and 121031300176-3.

\end{document}